\documentclass[twocolumn, amsmath,amssymb, aps, prl, preprint,10pt]{revtex4-1}

\usepackage{graphicx,color}
\usepackage[dvipsnames]{xcolor}
\usepackage{braket,hyperref,amsmath, cancel}

\newcommand\Ccancel[2][black]{
	\let\OldcancelColor\CancelColor
	\renewcommand\CancelColor{\color{#1}}
	\cancel{#2}
	\renewcommand\CancelColor{\OldcancelColor}
}

\newcommand\figref[1]{Fig.~\ref{#1}}
\newcommand\figsref[1]{Figs.~\ref{#1}}
\newcommand{\ip}{\mathrm{I}_\mathrm{p}}
\newcommand{\up}{\mathrm{U}_\mathrm{p}}
\newcommand{\pb}{\mathbf{p}}
\newcommand{\dd}{\mathrm{d}}
\newcommand{\rb}{\mathbf{r}}
\newcommand{\Ab}{\mathbf{A}}

\renewcommand{\deg}{^{\circ}}

\begin{document}
	\preprint{APS/123-QED}
	\title{Spiral-like Holographic Structures: Unwinding Interference Carpets of Coulomb-Distorted Orbits in Strong-Field Ionization}
	\author{Andrew S Maxwell}
	\email{andrew.maxwell@ucl.ac.uk}
	\author{Carla Figueira de Morisson Faria}
	\affiliation{Department of Physics \& Astronomy, University College London \\Gower Street  London  WC1E 6BT, United Kingdom}
	\author{XuanYang Lai}
	\author{RenPing Sun}
	\author{XiaoJun Liu}
	\affiliation{Wuhan Institute of Magnetic Resonance and Atomic and Molecular Physics, Wuhan Institute of Physics and Mathematics, Innovation Academy for Precision Measurement Science and Technology, Chinese Academy of Sciences, Wuhan 430071, China}
	\date{\today} 
	\begin{abstract}
		We unambiguously identify, in experiment and theory, a previously overlooked holographic interference pattern in strong-field ionization, dubbed ``the spiral", stemming from two trajectories for which the binding potential and the laser field are equally critical. We show that, due to strong interaction with the core, these trajectories are optimal tools for probing the target \textbf{after} ionization and for revealing obfuscated phases in the initial bound states. The spiral is shown to be responsible for interference carpets, formerly attributed to direct above-threshold ionization trajectories, and we show the carpet-interference condition is a general property due to the field symmetry.	
	\end{abstract}
	\pacs{32.80.Rm}
	\maketitle

The interaction of matter with intense laser fields ($I=10^{13}$~W/cm$^2$ or higher) has led to the inception of attoscience. Attoseconds are some of the shortest time scales in nature, which makes real-time probing and steering of electron dynamics possible \cite{Salieres2012R,Lepine2014,Leone2014,Lindroth2019}.   Examples are resolving charge migration \cite{Lepine2014,Calegari2014,Calegari2016} and conical intersections \cite{Conical2018} in molecules and using tailored fields to probe chiral  systems \cite{Ayuso2019,Rozen2019}. One must measure not only amplitudes, but also phase differences in order to reconstruct specific targets. This requirement  has caused the development of ultrafast photoelectron holography  \cite{Spanner2004, Bian2011, Huismans2011Science,Faria2019}, which combines two key advantages: high photoelectron currents and subfemtosecond resolution. It exploits the quantum interference of different paths that an electron can take during strong-field ionization to produce a holographic image of the target with all the important phase information. Typically, there is a direct (`\textit{reference}') pathway and one that re-interacts with the target (`\textit{probe}').
Useful phases imprinting 
structural information are thought to be acquired after ionization, when the `\emph{probe}' returns close by the parent ion.
Hence, for optimal imaging one should minimize the closest distance from the core upon return.
Examples of holographic patterns are a spider-like structure \cite{Huismans2011Science,Hickstein2012}, fan-shaped \cite{Rudenko2004b,Maharjan2006} and fishbone-type fringes \cite{Haertelt2016}.

The fan results from the interference of direct and lightly deflected trajectories \cite{Lai2017,Maxwell2017}, and the spider is caused by the interference of two types of forward-scattered trajectories, whose interaction
with the core is brief  \cite{Huismans2011Science,Hickstein2012,Maxwell2017}.   
Yet, they can be used to probe the target. Enhancements in the fan were related to the coupling of electronic and nuclear degrees of freedom in $H_2$ \cite{Mi2017}, while suppressions were associated with electron capture in Rydberg states in dimers \cite{Veltheim2013}. The spider has been shown to be sensitive to molecular orientation \cite{Meckel2014} and employed to investigate the dynamics of nodal planes and molecular alignment \cite{Walt2017}. Still, the above-stated effects either relate to the structureless Coulomb tail or to phase differences obtained \textit{prior} to ionization, such as those in the target's initial bound states. 
Nonetheless, there is a strong motivation to go beyond that scope, and image changes that happen subsequently to ionization, such as polarization, charge migration or multielectron dynamics. 
Thus, a stronger interaction with the core during continuum propagation is desirable.
An early example is the fishbone structure reported in  \cite{Haertelt2016}, which was associated with backscattered trajectories, but was obfuscated by the spider-like fringes. This made an elaborate scheme necessary in order to remove the spider artificially and retrieve such a structure. 
 
Alternatively, one may employ holographic patterns caused by backscattered trajectories that occur in momentum regions for which the spider is suppressed, such as interference carpets \cite{ Korneev2012,Korneev2012a,Becker2019,Kazemi2013a,Li2015,Almajid2017,Guo2017,Ren2019,Li2014a}. However, they were attributed to interfering direct SFA orbits and their interplay with  above-threshold ionization (ATI) rings \cite{Korneev2012a, Korneev2012}, but there is room for misinterpretation.   
First, the energy region for which they are observed is much higher than the direct ATI cutoff \footnote{The direct ATI cutoff occurs at a photoelectron energy of $2 \up$, where $\up$ is the ponderomotive energy. For a recent review see \cite{Becker2018}.}. 
Second, a theoretical study \cite{Li2015} of two colour orthogonally polarized fields concluded that the long-range Coulomb tail boosted the importance of forward-scattered trajectories in carpet-like interferences.
This invites the questions of why rescattering is not important and has not been studied in this context. 

It is noteworthy that most explanations of holographic structures neglect many important Coulomb effects.  This has led to Coulomb-distorted orbit-based methods \cite{Faria2019}, including the Coulomb quantum-orbit strong field approximation (CQSFA). The CQSFA enables an incredibly clear picture of quantum interference that has revealed a whole host of previously overlooked interference patterns \cite{Lai2015,Lai2017,Maxwell2017,Maxwell2017a,Maxwell2018}.  With this in mind we suggest that the explanation in \cite{Korneev2012a} for the carpet-like  structure is over simplified.

In this Letter, we provide a more general explanation and show that some properties of  the interference carpets may be caused by several types of interfering orbits, including rescattered ATI orbits and Coulomb distorted trajectories. In particular, in the high-energy photoelectron region, we find a spiral-like pattern resulting from the interference between Coulomb-distorted back- and forward-scattered electron trajectories, which is entirely responsible for the carpet-like structure. 
This spiral-like pattern is unambiguously verified in our experiments. Furthermore, we show the spiral is well suited to holographic imaging with a strong interaction with the core and has the ability to reveal usually obscured phases, thus providing a promising approach for ultrafast imaging of atomic and molecular structures.
	
The CQSFA \cite{Lai2015,Maxwell2017} takes the formally exact transition amplitude for single electron strong field ionization
\begin{equation}
	M(\pb)=-i \lim_{t\rightarrow \infty} \int_{-\infty }^{t }\dd
	t'\left\langle \psi_{\pb}(t)
	|\hat{U}(t,t')H_I(t')| \psi _0(t')\right\rangle \,
	,\label{eq:transitionampl}
\end{equation}
 where the initial state is taken as  $\left\vert \psi _{0} (t_0)\right\rangle= e^{iI_pt_0}\left\vert \psi
 _{0}\right\rangle$, $
 |\psi_{\textbf{p}_f}(t)\rangle$ is a final continuum state with momentum $\mathbf{p}_f$. The time evolution operator  $\hat{U}(t,t')$ relates to the full Hamiltonian $
 \hat{H}(t)=\hat{\mathbf{p}}^{2}/2+V(\hat{\mathbf{r}})+\hat{H}_I(t)$, where $V(\hat{\mathbf{r}})$ is the binding potential and the interaction with the field is given by $\hat{H}_I(t)=-\hat{\mathbf{r}}\cdot \mathbf{E}(t)$. Using a path-integral formalism coupled with the saddle-point approximation this becomes the sum 
\begin{equation}
	\label{eq:MpPathSaddle}
	M(\mathbf{p}_f)\propto-i \lim_{t\rightarrow \infty } \sum_{s}\bigg\{\det \bigg[  \frac{\partial\mathbf{p}_s(t)}{\partial \mathbf{r}_s(t_s)} \bigg] \bigg\}^{-1/2} \hspace*{-0.6cm}
	\mathcal{C}(t_s) e^{i
	S(\mathbf{p}_s,\textbf{r}_s,t,t_s))}
\end{equation}
over $s$ quantum orbits. The action along each orbit reads \begin{equation}\label{eq:stilde}
	S(\mathbf{p},\textbf{r},t,t')=\ip t'-\int^{t}_{t'}[
	\dot{\mathbf{p}}(\tau)\cdot \mathbf{r}(\tau)
	+H(\mathbf{r}(\tau),\mathbf{p}(\tau),\tau)]d\tau,
\end{equation}
where 
the intermediate momentum $\mathbf{p}$ and coordinate $\mathbf{r}$ have been parametrized in terms of the time $\tau$. 
The integral in Eq.~(\ref{eq:stilde}) diverges at the lower bound; this is fixed using the regularization procedure as described in \cite{Maxwell2018PRA,Popruzhenko2014,Popruzhenko2009}.
The variables $t_s$, $\mathbf{p}_s$ and $\mathbf{r}_s$ are determined by the saddle-point equations
\begin{eqnarray}\label{eq:s1} [
	\textbf{p}(t')+\textbf{A}(t')]^{2}/2+\ip=0 ,
\end{eqnarray}
\begin{eqnarray}\dot{\pb}(\tau)=
	-\nabla_\textbf{r}V[\rb(\tau)]
\ \text{ and } \
	\dot{\rb}(\tau)= \pb(\tau)+\Ab(\tau).
	\label{eq:s2} 
\end{eqnarray}
The term in brackets is associated with the stability of the orbit, and $\mathcal{C}(t_s)$ is given by
\begin{equation}
	\label{eq:Prefactor}
	\mathcal{C}(t_s)=\sqrt{\frac{2 \pi i}{\partial^{2}	S(\mathbf{p}_s,\textbf{r}_s,t,t_s) / \partial t^{2}_{s}}}\langle \mathbf{p}+\mathbf{A}(t_s)|H_I(t_s)|\Psi_{0}\rangle,
\end{equation}	
where $\ket{\Psi_{0}}$ refers to the initial bound state of the electron, which we find using the GAMESS-UK \cite{GAMESS-UK} quantum chemistry software. In this work we will use the ground states of xenon, neon and helium.

In contrast to all previous calculations \cite{Lai2015,Lai2017,Maxwell2017a,Maxwell2017a,Maxwell2018,Maxwell2018PRA,Kang2019}, which used the one-over-r Coulomb potential for the continuum phase and dynamics \footnote{For the tunnelling step the species is already fully accounted for thus an effective potential is not required.}, here we employ a single-electron effective potential for noble gases \cite{Milosevic2010,Tong2005}, which takes the form
\begin{equation}
	V(\rb(\tau))=-\frac{1+f(r(\tau))}{r(\tau)},
	\label{eq:eff_pot}
\end{equation}
where $f(r)=a_1 e^{-a_2 r}+a_3 r e^{-a_4 r}+a_5 e^{-a_6 r}$ and $r(\tau)=\sqrt{\rb(\tau)\cdot\rb(\tau)}$.
The $a_i$ parameters are set by fitting to a numerically calculated potential \cite{Tong2005}. For the atoms of interest the values are listed in Table \ref{tab:a_values}. We will denote a simple hydrogenic Coulomb, helium, neon and xenon potentials as $V_{H}$, $V_{He}$, $V_{Ne}$ and $V_{Xe}$, respectively. These potentials ensure that `structural' information is imprinted in the phases, and we can use this to evaluate the sensitivity of the holographic structures. Similar to previous work \cite{Maxwell2017} one may simplify the action using  
\begin{equation}
\dot{\pb}(\tau)\cdot\rb(\tau)=V(\rb(\tau))-f'(r(\tau)).
\end{equation}

\begin{table}
	\begin{tabular}{|l|l|l|l|l|l|l|}
		\hline
		Atom & $a_1$ & $a_2$& $a_3$& $a_4$& $a_5$& $a_6$\\
		\hline
		hydrogen & 0 & 0 & 0 & 0 & 0 & 0\\
		helium & 1.231 & 0.662 & -1.325 & 1.236 & -0.231 & 0.480\\
		neon & 8.069 & 2.148 & -3.570 & 1.986 & 0.931 & 0.602\\
		xenon & 51.356 & 2.112 & -99.927 & 3.737 & 1.644 & 0.431\\
		\hline
	\end{tabular}
\caption{The $a_i$ parameters for the single electron effective potential for Nobel gases. Values taken from \cite{Milosevic2010,Tong2005}.}
\label{tab:a_values}
\end{table}

\begin{figure}
	\includegraphics[width=\linewidth]{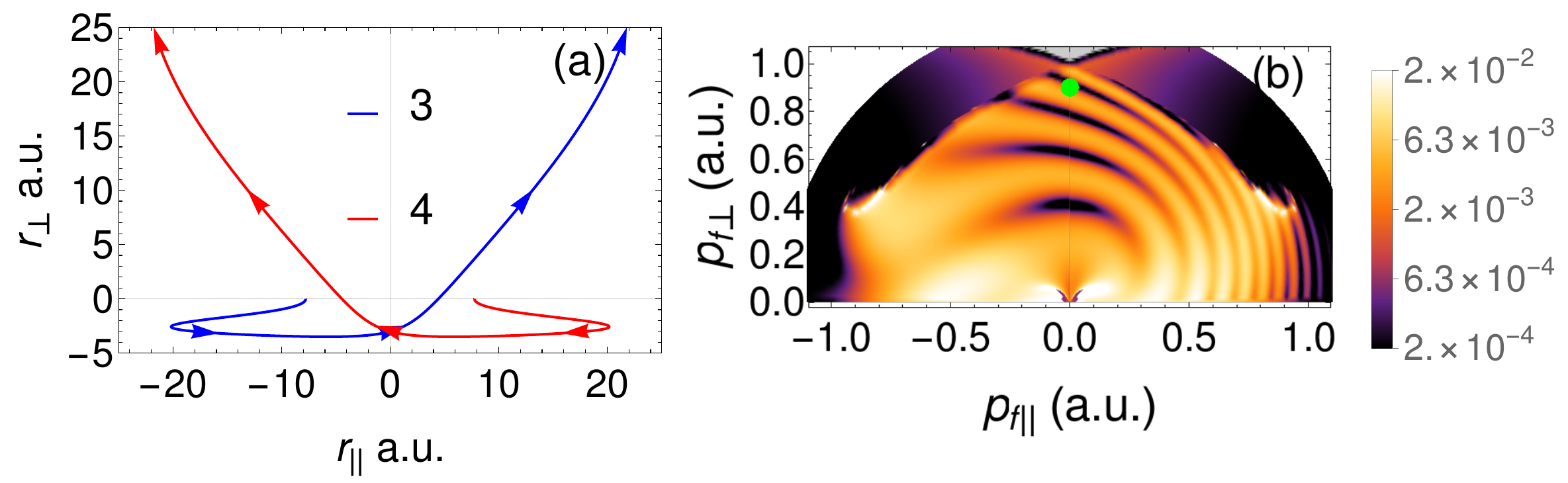}
	\caption{Panel (a) shows the two example trajectories for orbit 3 and 4 for final  low parallel and high transverse momentum components, marked by green spot on panel (b). Panel (b) shows the resulting spiral-like interference structure that occurs in the photoelectron momentum distribution. This was computed using the CQSFA for a laser intensity $I=7 \times 10^{13}$~W/cm$^2$ and wavelength of 800~nm over a single cycle for a xenon target with $\ip =0.446$~a.u., initial wave function calculated by GAMESS \cite{GAMESS-UK} and using the potential $V_{Xe}$ in the continuum.}
	\label{fig:Diagram}
\end{figure}

The classification, from \cite{Yan2010}, for the four orbits is as follows:
For orbit 1, the electron tunnels towards the detector and reaches it directly. If it is released on the ``wrong'' side, and then turns around to reach the detector, the electron will propagate along orbit 2 or 3. 
For orbit 2, the initial and final transverse momentum of the electron will point in the same direction, while for orbit 3 it will reverse. For orbit 4, the electron is freed towards the detector, but scatters off core. 
One may view orbit 1 as direct, orbits 2 and 3 as forward scattered and orbit 4 as backscattered.  
For details see our previous work \cite{Maxwell2018} and the review \cite{Faria2019}.
In \figref{fig:Diagram}(a) we show the CQSFA orbits 3 and 4, whose  interference gives spiral-shaped fringes, shown in panel (b) \cite{Maxwell2018}.  The spiral-shaped fringes are most clearly observed in the high-energy part of the distribution close to the perpendicular momentum axis.  

In order to measure the spiral in experiment, we employ a commercial laser system (FEMTOPOWER Compact PRO, Femtolasers Produktions GmbH) consisting of a broadband femtosecond oscillator and a multipass chirped-pulse amplifier. The system delivered 30-fs pulses (FWHM) with a maximal output energy of 0.8 mJ, a central wavelength of 800 nm, and a repetition rate of 5 kHz. The pulse energy from the amplifier was adjusted by means of a broadband achromatic half-wave plate followed by a thin-film polarizer. The linearly polarized pulses were focused into the interaction chamber by an on-axis spherical mirror with a focal length of 75 mm. The sample gas was fed into the interaction chamber through a needle valve. The ejected electrons were detected using a velocity map imaging (VMI) spectrometer \cite{Eppink1997}. Images were recorded using a delay-line position-sensitive detector. Retrieval of the velocity and angular distribution of the measured photoelectrons was performed by using the Gaussian basis-set expansion Abel transform method \cite{Dribinski2002}. 
	\begin{figure}
		\includegraphics[width=\linewidth]{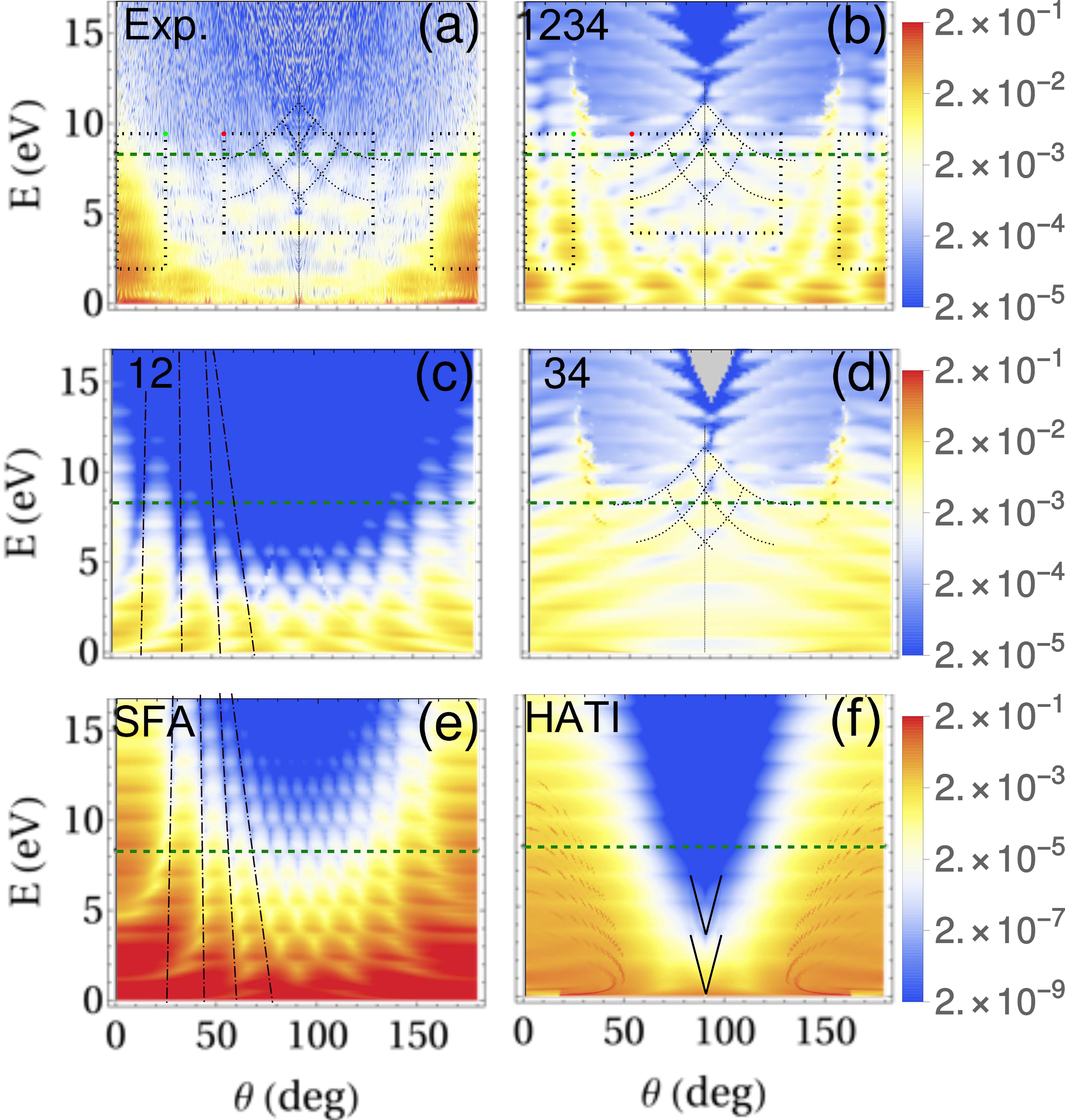}
		\caption{Photoelectron signal for xenon plotted for emission angle and energy.
			Panel (a): Experimental photoelectron signal for strong field ionization of xenon subject to an 11 cycle pulse of peak intensity $I=7\times10^{13}$~W/cm$^2$ and wavelength $\lambda=800$~nm ($\omega\approx1.55$~eV). Panel (b): Theoretical result employing the CQSFA for the same field parameters, including all electron orbits 1-4. This has been computed for 4 laser cycles. Panel (c): Same as previous, including only orbits 1 and 2. Panel (d): same as previous including only orbits 3 and 4.
			Panel (e): Theoretical results using the direct ATI orbits in the standard strong-field approximation following \cite{Korneev2012a}.	
			Panel (f): signal computed using high-order rescattered ATI (HATI) (see \cite{Lohr1997, Maxwell2018} for more details). Note the HATI prefactors have been neglected here as we are focusing on the interference fringes. 
			All theoretical results have been averaged over the focal volume of the laser field \cite{Kopold2002}. 
			The spiral-like fringes in the experimental results have been traced by black dotted lines, which have been duplicated to panels (b) and (d). The $2\up$ cut-off has been marked by a green dashed line. The central (outer) dotted rectangle(s) mark the region where the spiral (spider) is dominant and easily visible in experiment.  The fan and SFA fringes are marked by black dot-dashed lines. Initial states are computed using GAMESS UK \cite{GAMESS-UK} and as before using the potential $V_{Xe}$. A logarithmic scale is used over 4 orders of magnitude for panels (a)-(d) and 8 orders of magnitude for (e) and (f). The scale is in arbitrary units normalized with regard to the peak value in each panel.}
		\label{fig:MainResult}
	\end{figure}
	
In \figref{fig:MainResult}(a) we show  experimental results for strong-field ionization of xenon compared with CQSFA computations [\figref{fig:MainResult}(b)-(d)]. \figref{fig:MainResult} is plotted over the photoelectron emission angle ($\theta$) and energy ($E$), in order to distinguish the  spiral [V-shape] from ATI rings [horizontal lines]. The CQSFA results have been averaged over the focal volume according to \cite{Kopold2002}.
There is excellent agreement between the experiment and the CQSFA, panels (a) and (b), respectively, aside from a slight shift of $0.7$~eV. However, one may show that the polarizability of xenon will shift the ionization potential, and hence the carpet, by $0.75$~eV \cite{Spiewanowski2015}. 
In the high-energy region around $\theta\approx90\deg$ the V-shaped (spiral) and the horizontal (ATI rings) fringes combine to make oval shapes. In \figref{fig:MainResult}(d) the contributions of orbits 3 and 4 are plotted. Both the V-shaped structure and ovals are reproduced in the energy region of interest. If orbit 4 is removed the ovals disappear (see supplementary material), worsening the agreement with experiment.  Hence, the spiral is unambiguously identified as the cause of these high-energy fringes. The combination of the spiral-like structure and ATI rings in this energy region leads to the interference carpets.

 One of the main features is that, along the line $\theta=90\deg$, there is a spacing of $2\omega$ between the ovals obeying
 \begin{equation}
 \ip+\up+E_k=2n\omega,
 \label{eq:carpet}
 \end{equation}
 where n is an integer and $E_k=1/2p_{\perp}^2$ is the electron's kinetic energy \cite{Korneev2012,Korneev2012a}, which we also find in both theory and experiment; see \figref{fig:MainResult}. This gap stems from the mirror symmetry (due to the laser field) about the $r_{\perp}$ axis for pairs of interfering trajectories separated by exactly one half cycle, which leads to almost all phases cancelling out, including those related to the Coulomb potential. 
In the supplementary material we show analytically that Eq.~(\ref{eq:carpet}) is universal and is satisfied by pairs of direct and rescattered (high-order) ATI (HATI) SFA orbits and CQSFA trajectories. Like ATI rings, Eq.~(\ref{eq:carpet}) is due to a fundamental symmetry present for a linear monochromatic fields. In the CQSFA,  Eq.~(\ref{eq:carpet}) holds for the orbit pairs $(1,2)$ and $(3,4)$. Hence, demonstrating that a model satisfies Eq.~(\ref{eq:carpet}) is not sufficient evidence of the physical mechanism for the carpet.

There are fundamental discrepancies for the carpet-like structure between the present CQSFA interpretation and the direct SFA previously given in \cite{Korneev2012,Korneev2012a} [see Figs.~\ref{fig:MainResult}(b) vs (e)], namely: 
1) The signal associated with direct orbits is very low in the region of interest, while the contributions of orbits 3 and 4 dominate. 2) The fringes due to the interference of direct orbits lead to sharp V-shaped fringes, which are much finer than the chequerboard spiral/ carpet seen in experiment [Fig.~\ref{fig:MainResult}(a)].
 The figure shows that, although the carpet structure is reproduced at exactly $\theta=90^{\circ}$, there is significant disagreement away from $\theta=90^{\circ}$. The yield is strongly suppressed, and the V-shaped fringes are steeper and much finer. If one uses Coulomb-distorted (CQSFA) orbits 1 and 2 [Fig.~\ref{fig:MainResult}(c)] the fringes are slightly straighter, as Coulomb distortions lead to fan-shaped, nearly radial fringes \cite{Lai2015,Maxwell2017} in the momentum distributions. This worsens the agreement with experiment and refutes the explanation in \cite{Korneev2012a}, where very similar laser parameters were used for xenon.
In Fig.~\ref{fig:MainResult}(f), we plot contributions from two pairs of HATI orbits with ionization times separated by half a cycle (we use the uniform approximation in \cite{Faria2002}). 
However, in this case there is almost no signal in the region of interest and the interference washes out very rapidly away from $\theta=90\deg$. Thus, SFA type orbits fail to qualitatively reproduce the interference carpet. 

Three reasons make the spiral an ideal candidate for extracting information about the residual core via electron holography. 
Firstly, it is visible without any additional manipulation 
because in the angle-energy region of interest only electron orbits 3 and 4 are dominant allowing for a simpler analysis.  
Secondly, for $\theta=90\deg$,  phase differences that are usually hidden can be extracted \cite{Kang2019}.
Thirdly, these two trajectories  revisit the ion core very closely, undergoing the most interaction with the binding potential.

\begin{figure}
	\includegraphics[width=\linewidth]{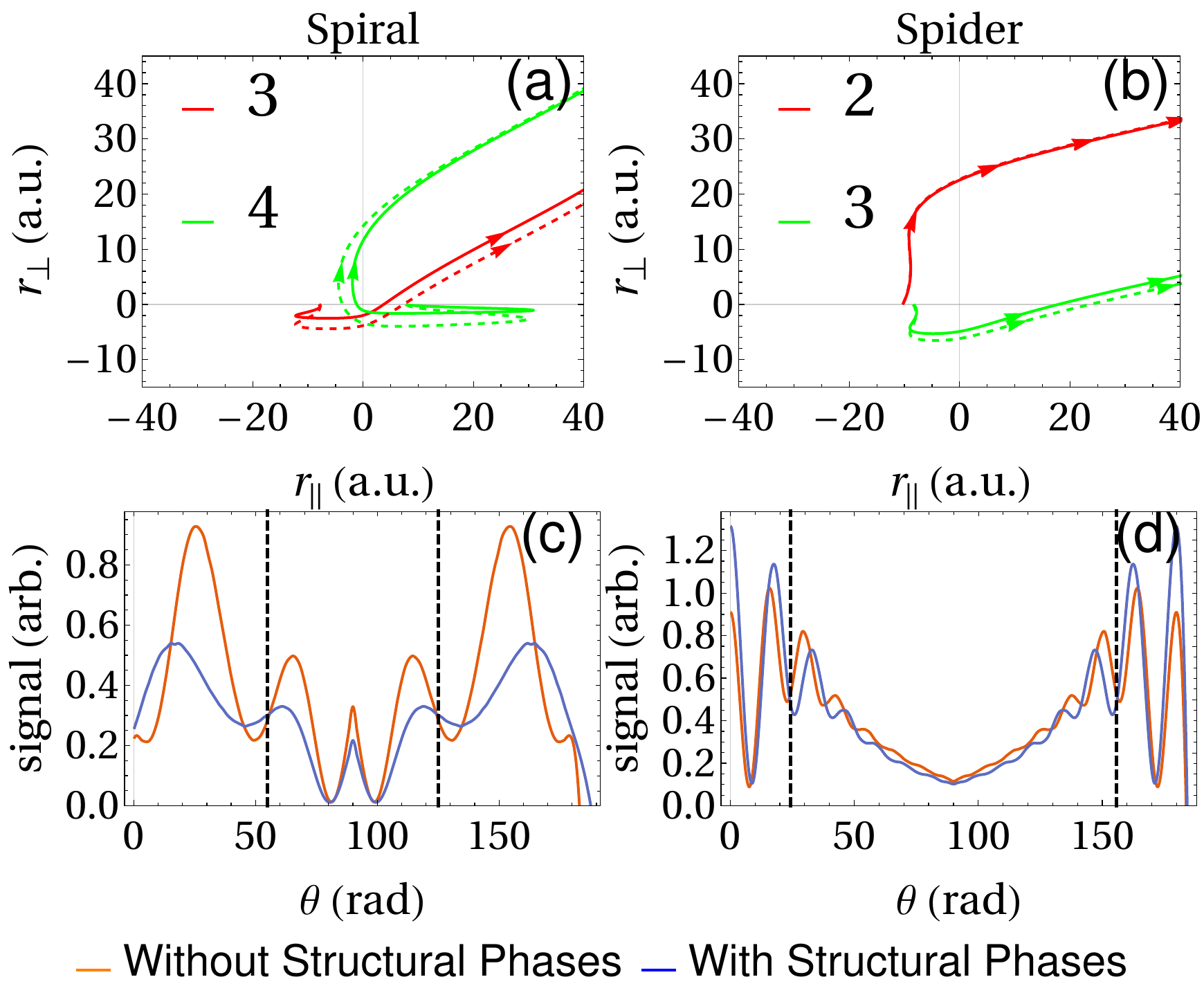}
	\caption{Top row: trajectories responsible for spiral, panel (a), and spider, panel (b), for the same target [xenon] and field parameters as Fig.~\ref{fig:MainResult}, with a final energy $E=9.5$~eV and angles $\theta=53 \deg$ and $\theta=24 \deg$ for panel (a) and (b), respectively.
	For the solid lines, a structureless Coulomb potential was used, while dashed line corresponds to trajectories calculated using the effective potential for xenon.
	Bottom row: CQSFA photoelectron signal for
	spiral ,panel (c) and spider, panel (d), for a fixed energy $E =9.5$~eV plotted over emission angle. No focal averaging has been used. Orange lines consider the structureless Coulomb potential in the continuum, while for blue lines the effective potential for xenon has been employed. Inside (outside) the black dashed line is the region where the spiral (spider) is dominant as in Fig.~\ref{fig:MainResult}. }
	\label{fig:FigCrossSec}
\end{figure}

This is exemplified in \figsref{fig:FigCrossSec}(a) and (b), where the contributing orbits for the spiral and spider are plotted. The final momentum of the orbits for each structure is chosen so that they interact the most with the binding potential. The orbits follow noticeably different paths for the spiral depending on whether $V_{H}$ or $V_{Xe}$ is used for continuum dynamics, while much less difference can be observed for the spider. Comparing the electron's closest distance to the core we find it is roughly twice as large for the spider [Table \ref{tab:Spiral_Comp} ]. At these distances $V_{Xe}$ and $V_{H}$ differ by 58\% and 6\% for the spiral and spider, respectively [see Table \ref{tab:Spiral_Comp} ]. Thus, the spiral is much more sensitive to the structural information encoded in the effective potential. In panels (c) and (d), we compare the spiral and the spider directly by plotting their signal as a function of the emission angle for a fixed energy $E=9.5$~eV.   The maximum deviation in position and height peak in the regions where spiral or spider dominate, given in Table \ref{tab:Spiral_Comp},  are about 2.5 and 6 times greater, respectively, for the spiral. This confirms a much stronger sensitivity to the target and structural phases.

\begin{table}
	\begin{tabular}{|l|p{1.8cm}|l|l|l|}
		\hline
		Pattern & closest\newline approach $r_c$ & $V_{Xe}(r_c)/V_H(r_c)$ & angle dev & peak dev\\
		\hline
		Spiral & 2.9~a.u. & 1.58 & $4.5\deg$ & 0.66\\
		Spider & 6.0~a.u. & 1.06 & $1.7\deg$ & 1.11\\ \hline
	\end{tabular}
	\caption{Comparison of the spiral and spider patterns sensitivity to structural phases encoded in the effective potential for xenon. The second column shows the closet point of either of the interfering trajectories given by $r_c$. Column 3 gives the ratio of the effective potential $V_{Xe}$ vs the Coulomb $V_{H}$. Column 4 and 5 give the deviation in position and height, respectively, of the peaks from Fig.~\ref{fig:FigCrossSec} panel (c) and (d).}
	\label{tab:Spiral_Comp}
\end{table}

\begin{figure}
	\includegraphics[width=\linewidth]{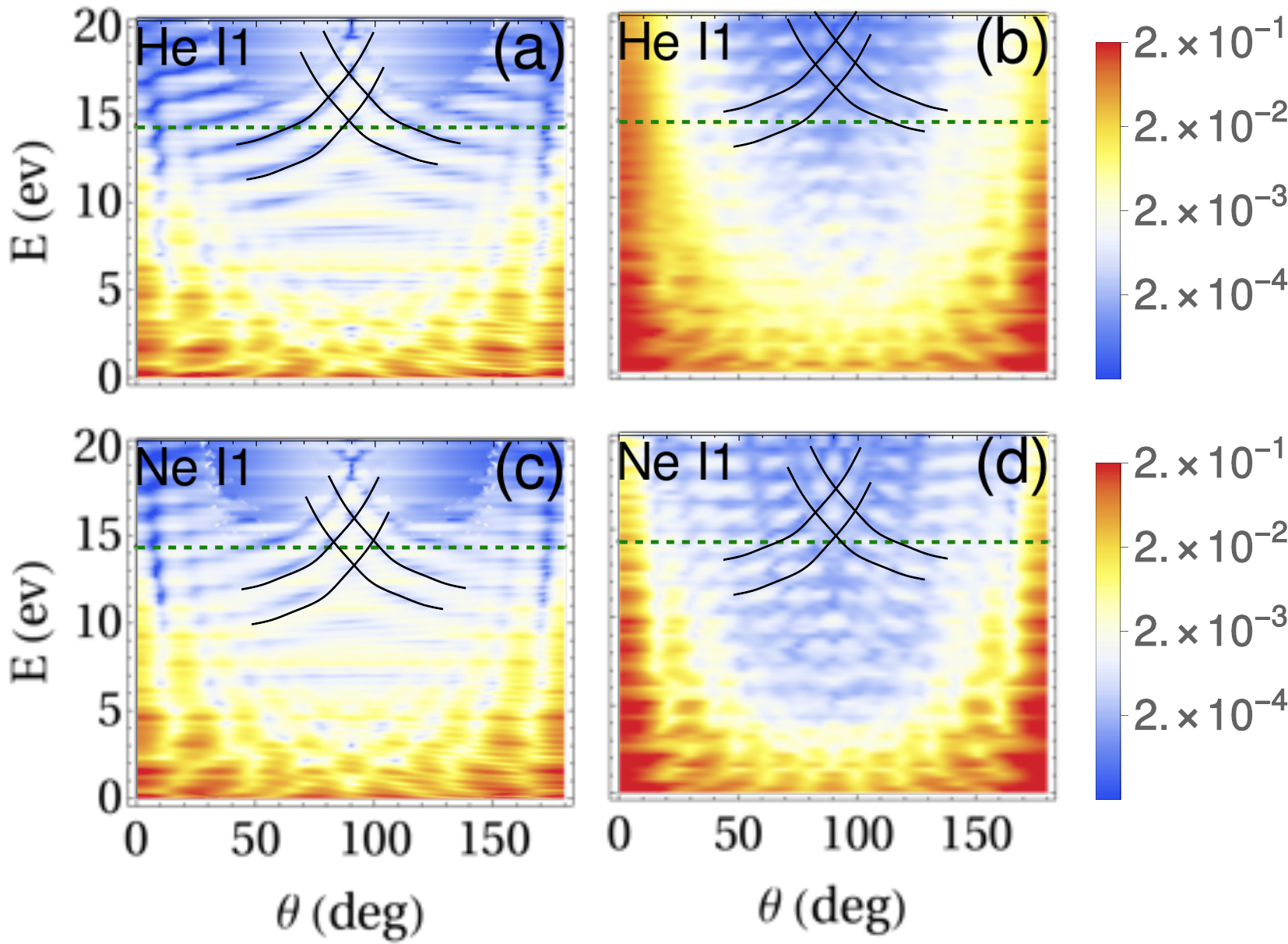}
	\caption{CQSFA [(a), (c)]] and QProp \cite{BauerQprop2006} calculations [(b), (d)] for helium [(a), (b)] and neon [(c), (d)] for the same frequency and number of laser cycles as \figref{fig:MainResult} with  an intensity $I=1.2 \times 10^{14}$W/cm$^2$. The results have been focally averaged \cite{Kopold2002}. The $2\up$ cut-off is marked by a green dashed line.
	}
	\label{fig:HeNePlots-20eV}
\end{figure}

In \figref{fig:HeNePlots-20eV} we compare CQSFA [panels (a) and (c)] and TDSE [panels (b) and (d)] calculations for helium and neon. The spiral like-fringes are visible in the highest energy region near $\theta=90^{\circ}$, but are less prominent than for xenon as they occur at a higher photoelectron energy. This demonstrates that the spiral-like interference is not limited to xenon or those particular laser parameters.
As before lines have been placed on the figures to trace the spiral fringes. The lines are shifted by a single photon energy ($\approx 1.4$ eV) between helium and neon. This is due to orbits 3 and 4 leaving from opposite side of the atom, and the valence orbitals of helium and neon having opposite [even and odd, respectively] parities. This leads to the two sets of fringes being out of phase. 
Note that the fringes between the CQSFA and Qprop calculation are shifted but crucially both models show out-of-phase fringes between targets. Comparing two targets, such as $N_2$ and neon in \cite{Kang2019}, allows the spiral to be exploited as a sensitive probe of orbital parity.

In conclusion, we have found a holographic spiral-like structure first predicted in \cite{Maxwell2018}, both in experiment and theory, and have identified it as the cause of interference carpets. We find that the 2$\omega$ gap in the interference carpets is a universal feature inherent to the field symmetry, which can be satisfied by many pairs of trajectories across different models for ATI. 

The spiral has been overlooked until now, for the following reasons. First, other explanations for the interference carpets \cite{Korneev2012,Korneev2012a,Guo2017, Almajid2017}, based on the direct SFA, were available and it was assumed that the 2$\omega$ gap was specific to that physical mechanism. Second, there was lack of proper theoretical treatment of orbits in that parameter range, due to the SFA being constructed as a field-dressed Born series using Coulomb-free trajectories. This left out a wide range of orbits considered by the CQSFA, for which both the potential \textit{and} the field were relevant. Thus, the prospects of the spiral for holographic imaging have not been realized. 

So far, interference carpets have solely been used for determining initial phases such as those associated with bound-state parity. Yet, the spiral is ideal for holographic imaging due to its high sensitivity to structural Coulomb phases. Furthermore, in contrast to the fishbone structure \cite{Haertelt2016}, it requires no further manipulation to be observed. Finally, the half-cycle separation between the pathways that form the spiral means that ultrafast dynamics could be resolved.  All of this makes the spiral the ideal structure for imaging and photoelectron holography. 

\textbf{Acknowledgements: } This research was in part funded by the National Key Research and Development Program of China (No. 2019YFA0307702), the National Natural Science Foundation of China (No. 11834015, No.11874392, No. 11922413), and the Strategic Priority Research Program of the Chinese Academy of Sciences (No. XDB21010400), and by the UK Engineering and Physical Sciences Research Council (EPSRC) (grants EP/J019143/1 and EP/P510270/1). The latter grant is within the remit of the InQuBATE Skills Hub for Quantum Systems Engineering. We thank the Wuhan Institute of Physics and Mathematics, Chinese Academy of Sciences, for its kind hospitality, and A. Staudte and H. Kang for useful discussions.

\bibliography{ExtractingSprial}{}
\bibliographystyle{prsty}
\end{document}